\documentclass[prd,nofootinbib,shownopacs,preprint]{revtex4}
\usepackage{graphicx}
\usepackage{epsfig}
\usepackage{amsmath}
\usepackage{amsfonts}
\usepackage{amssymb}
\usepackage{url}
\usepackage{hyperref}
\usepackage{subfigure}
\usepackage{hhline}
\usepackage{color}
\usepackage{bm}
\usepackage{cancel}
\usepackage{xspace}
\usepackage{blindtext}
\usepackage[italic]{hepnames}
\usepackage{amsmath,mathtools}
\usepackage{multirow}
\usepackage{booktabs}     

\newcommand{\be}{\begin{equation}}
\newcommand{\ee}{\end{equation}}
\newcommand{\bea}{\begin{eqnarray}}
\newcommand{\eea}{\end{eqnarray}}

\newcommand{\Bbar}{\,\overline{\!B}{}}
\newcommand{\Dbar}{\,\overline{\!D}{}}
\newcommand{\Kbar}{\,\overline{\!K}{}}
\def\B0bar{\Bbar{}^0}
\def\D0bar{\Dbar{}^0}
\def\K0bar{\Kbar{}^0}

\usepackage[toc,page]{appendix}
\usepackage{comment}
\begin{document}
\title{\bf Quantum decoherence signatures in charmless non-leptonic $B$ decays}

\author{Dhiren Panda}
\email{pandadhiren530@gmail.com}
\author{Manas Kumar Mohapatra}
\email{manasmohapatra12@gmail.com}
\author{Rukmani Mohanta}
\email{rmsp@uohyd.ac.in}

\affiliation{School of Physics,  University of Hyderabad, Hyderabad-500046,  India
}
\begin{abstract}
Quantum coherence plays a crucial role in the dynamics of neutral meson systems, aiding in the extraction of various Standard Model parameters. However, real physical systems always interact with their surroundings, which causes decoherence. In case of time dependent analysis of non-leptonic neutral $B$ meson decays, this decoherence can be modeled using a single parameter, $\lambda$. Since decoherence can affect the observed dynamics of flavor oscillations and CP violation, it becomes essential to revisit the key SM parameters such as the CKM angles $(\alpha, \beta, \gamma)$ and the mass differences of neutral $B$ mesons ($\Delta m_{d,s}$). In this work, we study the CKM phase  $\beta$ as well as the penguin contributions in the $B_d^0 \to J/\psi K_S$ decay mode in the presence of decoherence. We employ the pseudo-experiment (Toy Monte-Carlo) technique and perform an $SU(3)_F$ analysis using the $B_d^0 \to J/\psi \pi^0$ process. Furthermore, we investigate the $B_d^0 \to \pi^+ \pi^-$ decay mode to understand how the decoherence influences the CP violating observables. Our findings reveal that the presence of decoherence can affect crucially the measured values of the observables.
\end{abstract}

\maketitle

\section{Introduction}
The study of flavor physics, particularly the $B$ meson decays, offers a powerful tool to test the Standard Model (SM) and its possible extensions. Although the energy frontier experiments, in particular ATLAS and CMS, have not yet observed any direct evidence of new physics, precision measurements in the flavor sector provide a complementary approach in this direction. In this context, experiments such as LHCb and Belle II are uniquely sensitive to indirect signatures of new interactions that may arise at energy scales beyond the current collider reach. Moreover, the flavor observables can also probe the tiny scale effects associated with quantum decoherence and potential Lorentz symmetry violation~\cite{Mavromatos2005,ELLIS1992142,Huet:1994kr}. In this work, we specifically investigate how quantum decoherence may affect the $B$ meson system and influence the fundamental parameters.

This idea traces back to early theoretical insights, where Hawking’s study of black hole radiation revealed a connection between quantum mechanics and gravity, motivating the exploration of possible decoherence effects in particle systems. They performed the pure quantum states can turn into mixed states because of information loss across the event horizon \cite{Hawking:1982dj,Ellis:1983jz}. This phenomenon, known as quantum decoherence, implies a fundamental departure from unitarity and raises the possibility of CPT violation. The unique structure of spacetime at the Planck scale can lead to the spontaneous formation and evaporation of microscopic quantum black holes through Hawking radiation \cite{Hawking:1982dj}. 
Similar behavior arises in non-critical string theories and quantum gravity, where the discrete structure of spacetime naturally leads to decoherence \cite{Huet:1994kr}. Moreover, in the cosmological context, the accelerated expansion of the Universe toward a de Sitter phase introduces a cosmological horizon that can act as an additional source of decoherence \cite{Ellis:1983jz}. Regardless of its origin, whether from microscopic gravity effects or large-scale cosmological phenomena, the influence of the environment on a quantum system can be systematically described using the open quantum system formalism \cite{PhysRevE.67.056120,breuer2002theory,CABAN2007389}. The combined system and its environment evolve unitarily, preserving overall quantum coherence. However, when only the system is considered by tracing out the environment, its evolution becomes non-unitary and can be effectively described using the open quantum system formalism \cite{Lindblad:1975ef,Matthews:1961vr,Caban:2007je,Caban:2005ue}. This systematically accounts for decoherence and dissipation effects.

Motivated by the possibility that decoherence may influence precision measurements of the system, several studies have been conducted in $B$ meson decays. In Ref.~\cite{Alok:2024amd}, the authors investigated quantum decoherence in $B$ meson systems, including effects from decay width difference and CP violation in mixing. They identified that observables in leptonic, semileptonic, and non-leptonic $B$ decays are sensitive to decoherence and proposed that precision measurements from Belle~II and LHCb could constrain the decoherence parameter across multiple decay channels. Similarly, in Ref.~\cite{Alok:2025duw}, the authors performed the first combined analysis of mixing asymmetry and CP asymmetry measurements for $B_d$ mesons to constrain quantum decoherence parameters, and found the decoherence parameter $\lambda_d$ to be non-zero at about $6\sigma$ significance. They also set the first experimental limits for the decoherence parameter $\lambda_s$ in $B_s$ mesons, finding it to be non-zero at $3\sigma$ significance. Additionally, the time evolution of neutral meson systems, including decoherence effects with explicit Kraus operator representations that study the decay, mixing, and decoherence effects have been discussed in Ref. \cite{Alok:2015iua}. On the other hand, the analysis in Ref.  \cite{Banerjee:2014vga} discusses the quantum correlations in unstable meson-antimeson systems ($K$, $B_d$, and $B_s$) using semigroup formalism, showing that their quantum correlation measures differ significantly from those in stable systems due to the decay effects.

In this context, we focus on the non-leptonic $B_d^0 \to J/\psi K_S$,  $B_d^0 \to J/\psi \pi^0$, and $B_d^0 \to \pi^+ \pi^-$ processes to probe the impact of the decoherence parameter. The $B_d^0 \to J/\psi K_S$ decay is known as the golden channel for determining the weak phase $\beta$ (or the CKM phase $\phi_{1}$). It plays a key role in testing the CP violation predicted by the SM. The related mode $B_d^0 \to J/\psi \pi^0$ constrains CKM angles using SU(3) flavor symmetry. 
The channel $B_d^0 \to J/\psi \pi^0$ provides complementary information on CP violation and penguin contributions in the $B_d^0$ system. 
Moreover, this channel is sensitive to intermediate charm meson effects and helps refine our understanding of decay dynamics in the process. Hence, these decays together serve for high-precision flavor physics and new physics searches, making them ideal for studying the quantum decoherence effects on the observables. The authors in Ref. \cite{Barel:2020jvf} performed an analysis of the non-leptonic decay $B_d^0 \to J/\psi K^0$ incorporating doubly Cabibbo-suppressed penguin contributions using $SU(3)$ flavor symmetry to control theoretical uncertainties. They extracted the $B_d^0-\bar{B}_d^0$ mixing phases $\phi_d\equiv 2 \beta$ precisely and determined effective color suppression factors, providing important benchmarks for QCD decay dynamics and new physics searches. The decoherence effects in these modes are neglected, but understanding them is important for accurately interpreting CP asymmetry measurements and probing possible signs of new physics. However, in Ref.~\cite{Alok:2025duw}, the authors examined the $B_d^0 \to J/\psi K_S$ mode by considering decoherence but neglecting penguin contributions. On the other hand, the decay mode $B_d^0 \to \pi^+ \pi^-$ is an important non-leptonic channel in $B$~physics, providing a valuable probe of CP violation and the CKM angle $\alpha$ (or $\phi_2$). Although it is not among the primary golden modes such as $B_d^0 \to J/\psi K_S^0$, this decay serves as a complementary channel sensitive to direct CP violation and penguin contributions. 
The decay modes $B_d^0 \to J/\psi \pi^0$ and $B_d^0 \to \pi^+ \pi^-$ have been well studied for CP violation \cite{BaBar:2008kfx, Jung:2012mp, Ligeti:2025xbk, Fleischer:2018bld, Huber:2021cgk}, but their quantum decoherence effects have not been explored, unlike the golden mode $B_d^0 \to J/\psi K_S^0$. Therefore, we concentrate on the non-leptonic decay modes $B_d^0 \to J/\psi K_S$, $B_d^0 \to J/\psi \pi^0$, and $B_d^0 \to \pi^+ \pi^-$ to examine how the decoherence parameter affects these processes.
For this analysis, we employ the toy Monte Carlo simulations by generating and fitting pseudo-data to evaluate the impact of the decoherence parameter. 

The structure of this paper is outlined as follows. In Section \ref{nl}, we present the decoherence effect on the non-leptonic $B$ meson decays. Section \ref{sec:framework} deals with the nomenclature used for the physical constants and parameters in the presence of decoherence. Section \ref{Simulations} focuses on the benchmark modes and the simulations for $B_d^0\to J/\psi K_{S}$  and $B_d^0\to \pi^+ \pi^-$ processes. The results  
are discussed in section \ref{Results}. The findings and conclusions are summarized in section \ref{conclusion}.

\section{Decoherence in non leptonic decays of $B$ mesons} \label{nl}

In this section, we examine the decay and oscillation dynamics of the $B$ meson system. 
The time evolution of these processes is described in a basis consisting of three states: 
$|B^0\rangle$, $|\bar{B}^0\rangle$, and $|0\rangle$, 
where $|0\rangle$ denotes the vacuum state (corresponding to the absence of a $B$ meson). 
Within this framework, the initial density matrix $\rho_{B^0(\bar{B}^0)}(0)$ represents a pure system 
that begins as either a $B^0$ or $\bar{B}^0$ state. 
The subsequent time evolution of these pure states is governed by a set of Kraus operators $K_i(t)$, 
and can be expressed as
\[
\rho(t) = \sum_i K_i(t)\, \rho(0)\, K_i^\dagger(t)\;,
\]
following the formalism described in Ref.~\cite{alicki2007quantum}. These Kraus operators are designed to account for the decoherence that arises due to the systems interacting with the environment \cite{Sudarshan1961920,Alok:2024amd}. The states associated with $B^0$ and $\bar{B}^0$ are given as

\bea
\rho_{B^0}(t)&=& \frac{1}{2} e^{-  \Gamma t}
\left(\begin{array}{ccc} 
 a_{ch} + e^{-  \lambda t} a_c &  (\frac{q}{p})^*(- a_{sh} -i  e^{-  \lambda t} a_s) & 0  \\ 
(\frac{q}{p})(- a_{sh} + i e^{-  \lambda t} a_s) & |\frac{q}{p}|^2 a_{ch} - e^{-  \lambda t} a_c  & 0 \\ 
0 &0   & 2( e^{-  \Gamma t}-a_{ch})
\end{array}\right),
\nonumber\\
\rho_{\bar{B}^0}(t)&=& \frac{1}{2} e^{-  \Gamma t}
\left(\begin{array}{ccc} 
|\frac{p}{q}|^2 (a_{ch} - e^{-  \lambda t} a_c) & (\frac{p}{q})(- a_{sh} +i  e^{-  \lambda t} a_s) & 0  \\ 
(\frac{p}{q})^* (- a_{sh} - i e^{- \lambda t} a_s) & a_{ch} + e^{-  \lambda t} a_c  & 0 \\ 
0 &0   & 2( e^{-  \Gamma t}-a_{ch})
\end{array}\right),
\label{dm-bbbar}
\eea
where $a_{ch}=\cosh \left(\frac{ \Delta \Gamma t}{2}\right)$, $a_{sh}=\sinh \left(\frac{ \Delta \Gamma t}{2}\right)$,
$a_c=\cos\left(\Delta m t\right)$ and $a_s=\sin\left(\Delta m t\right)$. $\Gamma=(\Gamma_L + \Gamma_H)/2$, $\Delta \Gamma = 
\Gamma_L -\Gamma_H$, with $\Gamma_L$ and $\Gamma_H$ being the decay widths of $B^0_L = p B^0 + q \bar B^0$ and $B^0_H = p B^0 - q\bar B^0$, respectively for light and heavy states of $B$ mesons. Here $p$ and $q$ are the complex coefficients that satisfy the condition $|p|^2+|q|^2 = 1$. Further, $\Delta m = m_H - m_L$, $m_L$ and $m_H$ are the masses of $B^0_L$ and $B^0_H$ states, respectively. 
Here, $\lambda$ represents the decoherence parameter that arises from the interaction between the one-particle system and its environment. Our primary motivation is to investigate the penguin contribution and the CKM phase $\beta$ using the decay processes $B_d^0 \to J/\psi K_S$ together with its U-spin partner $B_d^0 \to J/\psi \pi^0$. 
We further extend our analysis to the decay mode $B_d^0 \to \pi^+ \pi^-$ to investigate the possible impact of quantum decoherence.  Assuming negligible CP violation in mixing, we analyze the decays of $B_d^{0}/\bar{B}_d^{0}$ into CP eigenstates $f_{CP}$. The corresponding operators for these decays are given as follows \cite{Alok:2024amd}
\be 
\mathcal{O}_{f_{CP}} = |A_{f}|^2\left(\begin{array}{ccc} 
1 & (\frac{p}{q})\,\lambda_{f} & 0  \\ (\frac{p}{q})^*\lambda_{f}^*  &|\frac{p}{q}|^2 |\lambda_f|^2 &0 \\ 0 & 0 & 0
\end{array}\right).
\label{opfcp}
\ee
Here, the phase invariant quantity $\lambda_f$ is defined as 
\be 
\lambda_f = \frac{q}{p}\frac{\bar{A_f}\left(\equiv A(\bar B^0 \to f_{CP})\right)}{A_f \left(\equiv A(B^0 \to f_{CP})\right)}\,.
\ee
One can then define the $ CP$ violating observable for $B^0/ \bar B^0 \to f_{CP}$ as
\be 
{\mathcal{A}}_{f_{CP}} (t) = \frac{P_{f_{CP}}(B^0;t)-P_{f_{CP}}(\bar{B^0};t)}
{P_{f_{CP}}(B^0;t)+P_{f_{CP}}(\bar{B^0};t)},
\label{nonlep2}
\ee
where  the probability rates $P_{f_{CP}}(B^0;t)$ and $P_{f_{CP}}(\bar{B^0};t)$, including time dependence and decay width differences, are given as

\bea
\frac{P_{f_{CP}}(B^0;t)}{\frac{1}{2} e^{-  \Gamma t}|A_f|^2} &=&
\left(1+|\lambda_f|^2\right) \cosh \left(\frac{ \Delta \Gamma t}{2}\right) 
+\left(1-|\lambda_f|^2\right)e^{-  \lambda t} \cos\left(\Delta m t\right) 
\nonumber\\
&&- 2{\rm Re}(\lambda_f)\sinh \left(\frac{ \Delta \Gamma t}{2}\right)
-2 {\rm Im}(\lambda_f)e^{-  \lambda t} \sin\left(\Delta m t\right)\,,\nonumber\\
\frac{P_{f_{CP}}(\bar B^0;t)}{\frac{1}{2} e^{-  \Gamma t}|A_f|^2 |\frac{p}{q}|^2} &=&
\left(1+|\lambda_f|^2\right) \cosh \left(\frac{ \Delta \Gamma t}{2}\right) 
-\left(1-|\lambda_f|^2\right)e^{-  \lambda t} \cos\left(\Delta m t\right) 
\nonumber\\
&&- 2{\rm Re}(\lambda_f)\sinh \left(\frac{ \Delta \Gamma t}{2}\right)
+2 {\rm Im}(\lambda_f)e^{- \lambda t } \sin\left(\Delta m t\right)\,.
\label{nonlep1}
\eea
Thus, we have
\be
{\mathcal{A}}_{f_{CP}} (t) = \frac{e^{-\lambda t}\left[A_{\rm CP}^{\rm dir,\,f_{CP}}   \cos\left(\Delta m t\right) + A_{\rm CP}^{{\rm mix},\,f_{CP}} \sin\left(\Delta m t\right) \right]-\delta_B\left[\cosh \left(\frac{ \Delta \Gamma t}{2}\right)-A_{\Delta \Gamma}^{f_{CP}}  \sinh \left(\frac{ \Delta \Gamma t}{2}\right)\right]}{\cosh \left(\frac{ \Delta \Gamma t}{2}\right)+A_{\Delta \Gamma}^{f_{CP}}  \sinh \left(\frac{ \Delta \Gamma t}{2}\right)- \delta_B e^{-\lambda t}\left[A_{\rm CP}^{\rm dir,\,f_{CP}}   \cos\left(\Delta m t\right) + A_{\rm CP}^{{\rm mix},\,f_{CP}}  \sin\left(\Delta m t\right) \right]}\,,
\ee
where  $\delta_B$ is defined as 
\be
\delta_B \equiv \left (1 - \left|\frac{q}{p}\right|^2 \right ) \Big /\left (1 + \left|\frac{q}{p}\right|^2\right ),
\ee and
\begin{equation}
    A_{\rm CP}^{\rm dir,\,f_{CP}} = \frac{1-|\lambda_{f_{CP}}|^2}{1+|\lambda_{f_{CP}}|^2},  \quad A_{\Delta \Gamma}^{f_{CP}} = - \frac{2 {\rm Re}(\lambda_{f_{CP}})}{1+|\lambda_{f_{CP}}|^2}, \quad A_{\rm CP}^{{\rm mix},\,f_{CP}} = -\frac{2 {\rm Im}(\lambda_{f_{CP}})}{1+|\lambda_{f_{CP}}|^2}\,.
\end{equation}
With $\lambda=0$, we can obtain the usual expression of $CP$ asymmetry in the interference of mixing and decay \cite{WALDI20011}. However, the presence of decoherence certainly modifies the expressions of CP asymmetry in the present of mixing and decay.
 Neglecting $CP$ violation in mixing, we get a simplified expression as
\be 
{\mathcal{A}}_{f_{CP}} (t) = 
\frac{A_{\rm CP}^{\rm dir,\,f_{CP}}  \cos\left(\Delta m t\right)+ A_{\rm CP}^{{\rm mix},\,f_{CP}} \sin\left(\Delta m t\right)}
{ \cosh \left(\frac{ \Delta \Gamma t}{2}\right)+A_{\Delta \Gamma}^{f_{CP}}  \sinh \left(\frac{ \Delta \Gamma t}{2}\right)} e^{-  \lambda t} \,.
\label{cpasym1}
\ee
 When $\Delta \Gamma=0$, we get 
 \begin{equation}
\mathcal{A}_{f_{CP}} (t) \approx  \Big[A_{\rm CP}^{\rm dir,\,f_{CP}}  \cos\left(\Delta m t\right) +A_{\rm CP}^{{\rm mix},\,f_{CP}} \sin\left(\Delta m t\right)\Big]e^{-\lambda t}.
\end{equation}
Utilizing the above time-dependent analysis of two-body decays, one can obtain bounds on the decoherence parameter. Recent experimental results have reported large uncertainties in direct CP asymmetry measurements \cite{PhysRevD.107.052008}, suggesting a possible influence of penguin pollution in these decays. 
Moreover, since decoherence can obscure the time-dependent CP asymmetry, it effectively alters the contribution of penguin amplitudes. This, in turn, affects the extraction of the CKM parameters. Hence, it is important to reanalyze the data 
by including the effects of decoherence. Such an analysis allows for a better understanding of the penguin parameters and provides more reliable bounds on the decoherence parameter.

\section{Nomenclature}\label{sec:framework}
In this section, we present a framework to study penguin contributions in neutral $B_d^0$ meson decays to CP eigenstates. The analysis is carried out using toy Monte Carlo pseudo-experimental data combined with flavor symmetry with the effects of decoherence. Thus, the presence of decoherence modifies both $A_{\rm CP}^{\rm dir,\,f_{CP}}$ and $A_{\rm CP}^{{\rm mix},\,f_{CP}}$, thereby influencing the penguin contributions and the CKM phase.
We use this approach to the key decay channels such as  $B_d^0 \to J/\psi K_S^0$ and its U-spin partner $B_d^0 \to J/\psi \pi^0$. Additionally, we also explore the $B_d^0 \to \pi^+ \pi^-$ decay channel as well, and also the effect on the effective weak phase $\alpha_{eff}$ \cite{Zupan:2004hv}. 
To control hadronic uncertainty due to penguin topologies, one can define the decay amplitudes of neutral $B$ meson transitions in a unified framework. To establish the formalism, the starting point of the generic decomposition of amplitude is \cite{PhysRevD.60.073008} 
\begin{equation}
A(B_q^0 \to f) = \lambda_c^{(q)}\left[A_{cc}^f + A_{\rm pen}^{(c)f}\right]
+ \lambda_u^{(q)}A_{\rm pen}^{(u)f}
+ \lambda_t^{(q)}A_{\rm pen}^{(t)f},
\end{equation}
 where $A_{cc}^f$ denotes the current-current contributions and $A_{\rm pen}^{(q)f}$ describes the penguin  topologies with internal quark ($q=u,c,t$),
 where
\begin{equation}
\lambda_q^{(d)} \equiv V_{qd} V_{qb}^* ,
\end{equation} are the CKM factors.
Using Wolfenstein parametrization and CKM unitarity \cite{Wolfenstein:1983yz}, the above expression can be written in a compact form \cite{PhysRevD.60.073008,Fleischer:1999zi}
\begin{align}
    A(B_d^0\to f) & \equiv \phantom{\eta_f}\mathcal{N}_f\left[1-a_f e^{i\theta_f}e^{+i\gamma}\right]\:, \label{eq:TransAmp} \\
    A(\bar B_d^0\to f) & \equiv \eta_f\mathcal{N}_f\left[1-a_f e^{i\theta_f}e^{-i\gamma}\right]\:,
\end{align}
where $\mathcal{N}_f$ is the CP conserving normalization factor arising from the leading tree level contribution, $\eta_f$ is the CP-eigenvalue of the final state $f$, \( a_f \) denotes the relative magnitude of the penguin to tree contribution, \( \theta_f \) is the CP-conserving strong phase difference between tree and penguin amplitudes, and $\gamma$ is the weak phase. In this study, we fix the $\gamma$ value from HFLAV \cite{PhysRevD.107.052008}. We then extract the penguin parameters $a_f$ and $\theta_f$ by comparing CP asymmetries of U-spin related decay modes $B_d^0 \to J/\psi K_S^0$ and $B_d^0 \to J/\psi \pi^0$.
The time dependent CP asymmetry for the $B_d^0\to f$ decay without decoherence  is defined as \cite{Fleischer:1999nz,Fleischer:1999zi}
\begin{align}
    A_{\text{CP}}(t) & \equiv
    \frac{|A(B_d^0(t)\to f)|^2-|A(\bar B_d^0(t)\to f)|^2}{|A(B_d^0(t)\to f)|^2+|A(\bar B_d^0(t)\to f)|^2} \nonumber\\
    & = \frac{\mathcal{A}_{\text{CP}}^{\text{dir}}\cos(\Delta m_dt)+\mathcal{A}_{\text{CP}}^{\text{mix}}\sin(\Delta m_dt)}{\cosh(\Delta\Gamma_dt/2)+\mathcal{A}_{\Delta\Gamma}\sinh(\Delta\Gamma_dt/2)}\:,
\end{align} 
where $\Delta m_d\equiv m^{(d)}_{\text{H}}-m^{(d)}_{\text{L}}$ and $\Delta\Gamma_d\equiv \Gamma_{\text{L}}^{(d)}-\Gamma_{\text{H}}^{(d)}$ are the mass and decay width difference between the heavy and light mass eigenstates of the $B_d$ meson system, respectively.
Here the mass eigenstate rate asymmetry $\mathcal{A}_{\Delta\Gamma}$, the direct CP asymmetry $\mathcal{A}_{\text{CP}}^{\text{dir}}$ and the mixing-induced CP asymmetry $\mathcal{A}_{\text{CP}}^{\text{mix}}$ satisfy the relation
\begin{equation}
    \left[\mathcal{A}_{\text{CP}}^{\text{dir}}(B_q\to f)\right]^2 +
    \left[\mathcal{A}_{\text{CP}}^{\text{mix}}(B_q\to f)\right]^2  +
    \left[\mathcal{A}_{\Delta\Gamma}(B_q\to f)\right]^2 = 1\:.
\end{equation}
The CP asymmetries associated with the penguin parameters $a_f$ and $\theta_f$, and the $B_d^0$--$\bar B_d^0$ mixing phase $\phi_d$ as follows \cite{Fleischer:1999nz,Fleischer:1999zi}
\begin{align}
    \mathcal{A}_{\text{CP}}^{\text{dir}}(B_d^0\to f) & = \frac{2 a_f \sin\theta_f\sin\gamma}{1-2a_f\cos\theta_f\cos\gamma+a_f^2}\:, \label{eq:Adir}\\
    \eta_f\mathcal{A}_{\text{CP}}^{\text{mix}}(B_d^0\to f) & = \left[ \frac{\sin\phi_d-2 a_f \cos\theta_f\sin(\phi_d+\gamma)+a_f^2\sin(\phi_d+2\gamma)}{1 - 2a_f\cos\theta_f\cos\gamma+a_f^2}\right]\:.\label{eq:Amix}
\end{align}
If we consider only the leading tree-level contributions and neglect the small penguin effects, the parameter $a_f$ becomes zero ($a_f = 0$). In this case, the CP asymmetries take the simple forms as
\begin{equation}
    \mathcal{A}_{\text{CP}}^{\text{dir}} = 0, \qquad
    \eta_f\,\mathcal{A}_{\text{CP}}^{\text{mix}} = \sin\phi_d.
\end{equation}
This implies that there is no direct CP violation, and the mixing-induced CP asymmetry is determined solely by the weak phase $\phi_d$. Although various studies \cite{Barel:2020jvf,Fleischer:1999nz,Faller:2008zc} indicate that $a_f$ is small, the increasing experimental precision in the measurement of CP asymmetries for $B_d^0 \to J/\psi K^0$ means that this small contribution can no longer be neglected.
Hence, the determination of $\phi_d$ from the CP asymmetries in Eqs.~\eqref{eq:Adir} and \eqref{eq:Amix} requires knowledge of $a_f$ and $\theta_f$, together with the decoherence parameter $\lambda$. 
The $SU(3)$ flavour symmetry of QCD allows us to relate the hadronic parameters of the $\bar b \to \bar s c \bar c$ and $\bar b \to \bar d c \bar c$ transitions to one another. 
In this work, we consider the $B_d^0 \to J/\psi K^0$ decay together with its $U$-spin partner, $B_d^0 \to J/\psi \pi^0$ decay mode.
The expressions for the transition amplitudes and CP observables of the $B_d^0 \to J/\psi K^0$ mode are obtained by substituting \cite{Fleischer:1999nz,Fleischer:1999zi}
\begin{equation}\label{eq:sub_ccs}
    \mathcal{N}_f \to \left(1-\frac{\xi^2}{2}\right)\mathcal{A}'\:, \qquad
    a_f e^{i\theta_f} \to -\epsilon b' e^{i\rho'}\;,
\end{equation}
where $\mathcal{A}'$ is the hadronic amplitude \cite{Barel:2020jvf} and 
\begin{equation}
    \epsilon \equiv \frac{\xi^2}{1-\xi^2} = 0.05237 \pm 0.00027\:.
\end{equation}
The numerical value is calculated using $\xi \equiv |V_{us}|$ taken from Ref.\ \cite{ParticleDataGroup:2024cfk}.
$b'$ and $\rho'$ are the size and complex phase of the ratio between the hadronic amplitude of the penguin and tree decay topologies \cite{Barel:2020jvf}.
Together, $b'$ and $\rho'$ will be referred to as the penguin parameters.
On the other hand, the decay topologies in the control mode $B_d^0 \to J/\psi \pi^0$  mediated by $\bar b \to \bar d c \bar c$ quark level transitions exhibit a different dependence on the CKM matrix elements compared to those in $B_d^0 \to J/\psi K_{\text{S}}^0$.
For the process $B_d^0 \to J/\psi \pi^0$, the equivalent substitutions to Eq.\ \eqref{eq:sub_ccs} are \cite{Fleischer:1999nz,Fleischer:1999zi}
\begin{equation}\label{eq:sub_ccd}
    \mathcal{N}_f \to -\xi \mathcal{A}\:, \qquad
    a_f e^{i\theta_f} \to  be^{i\rho}\:.
\end{equation}
Comparing these to Eq.\ \eqref{eq:sub_ccs},  it should be noted that the penguin topology contribution does not include the factor $\epsilon$ hence the penguin effect is enhanced in the case of $\bar b\to \bar d c\bar c$ transition as compared to $\bar b\to \bar sc \bar c$ transition. However, the overall amplitude is suppressed by $\xi$, leading to a reduced decay rate and making these decays experimentally more difficult to study.

In the decay channels $B_d^0\to J/\psi K_{\text{S}}^0$ and $B_d^0\to J/\psi \pi^0$, the $U$-spin symmetry allows us to relate the penguin parameters to one another by the relation \cite{Barel:2020jvf}
\begin{equation}\label{eq:SU3_pen}
    b'e^{i\rho'} = be^{i\rho}\:,
\end{equation}
as well as the hadronic amplitudes \cite{Barel:2020jvf}
\begin{equation}\label{eq:SU3_had}
    \mathcal{A}'=\mathcal{A}\:.
\end{equation}
Now, the CP asymmetries in Eqs.\ \eqref{eq:Adir} and \eqref{eq:Amix}  depend only on these penguin parameters $a_{f}$ and $\theta_{f}$ 
which are the ratios of amplitudes, hence insensitive to hadronic uncertainties.
Similarly, the Eq.\ \eqref{eq:SU3_had} is liable to both factorizable and non-factorizable $SU(3)$ symmetry breaking. In contrast, Eq.\ \eqref{eq:SU3_pen}, receives solely the contribution from non-factorizable $SU(3)$ symmetry breaking since the dominant factorizable $SU(3)$ symmetry breaking correction cancels out in the ratio.
Therefore, the relation in Eq.,\eqref{eq:SU3_had} can serve as a useful test of $SU(3)$ flavor symmetry, as demonstrated in Refs. \cite{Bel:2015wha,Davies:2023arm,Belle-II:2024hqw}. Furthermore, this relation can be extended to the control channel $B_d^0 \to J/\psi \pi^0$ by neglecting contributions from exchange and penguin-annihilation topologies.
\section{Benchmark modes and the simulation}
\subsection*{CP violation via $B_d^0\to J/\psi K_{S}$} \label{Simulations}
This decay mode is regarded as the cleanest channel for determining the CKM angle $\beta$, owing to its negligible penguin contribution. However, recent experimental advancements \cite{PhysRevD.107.052008} indicate that, due to large statistical uncertainties, it can no longer be considered free from penguin effects. In this work, we study this decay mode and its  control channel under the effects of decoherence to constrain the penguin parameters and the CKM phase $\phi_d$. The control mode considered here is $B_d^0 \to J/\psi \pi^0$.
\subsection*{CP violation via $B_d^{0}\to \pi^{+} \pi^{-}$}
The decay $B_d^0 \to \pi^+ \pi^-$ receives contributions from both tree ($T$) and penguin ($P$) amplitudes \cite{Huber:2021cgk} arising from the quark level transition $\bar{b} \to \bar{u}u\bar{d}$. In the limit where the penguin contribution is negligible ($P \ll T$), the mixing-induced CP asymmetry reduces to $A_{\text{mix}} = \sin(2\alpha)$, with $\alpha$ being one of the CKM unitarity triangle angles. However, when penguin effects are significant, they change the observed asymmetry. In this case, the mixing-induced CP asymmetry becomes $A_{\text{mix}} = \sqrt{1 - A_{\text{dir}}^2}\sin(2\alpha_{\text{eff}})$ \cite{Zupan:2004hv}, where $A_{\text{dir}}$ is the direct CP asymmetry and $\alpha_{\text{eff}}$ represents the effective weak phase. The CKM phase is then related by $\alpha = \alpha_{\text{eff}} + \Delta\alpha$, where $\Delta\alpha$ represents the shift induced by penguin pollution. This correction can be obtained through an isospin analysis involving the decay modes $B_d^0 \to \pi^0 \pi^0$ and $B^+ \to \pi^+ \pi^0$.


\section*{SIMULATION}
We perform a toy Monte Carlo study to investigate the extraction of the CP asymmetry parameters 
$A_{\mathrm{dir}}$, $A_{\mathrm{mix}}$, the $B^0$--$\bar{B}^0$ mixing mass difference $\Delta m_d$, and the decoherence parameter $\lambda$. The goal of this work is to study the statistical sensitivities and evaluate the impact of decoherence on these observables. To validate our simulation method, we first reproduce the experimental results for the direct and mixing-induced CP asymmetries without including decoherence. We ascertain that the values are consistent with the experimental measurements. Then we model the function with decoherence. The results of the simulated data are illustrated in the tables, and their distributions are shown in the figures in the following subsections. For clarity, we provide detailed descriptions of the simulations and the statistical analysis below.

\subsection{Theoretical Framework}
A meson is produced in a pure flavour state $B^0/\bar B^0$ evolves in time according to the usual two-state formalism in neutral meson mixing. The oscillation probability is governed by the mass difference $\Delta m_d$ between the heavy $B_H$  and light $B_L$ mass eigenstates. For convenience, we again quoted the time-dependent asymmetry that constitutes the basis of the simulation. In the absence of decoherence, the time-dependent CP asymmetry for decays into a common   CP eigenstate $f_{CP}$ is: 
\begin{align}
A(t) &= A_{\mathrm{dir}} \cos(\Delta m_d \, t) 
     + A_{\mathrm{mix}} \sin(\Delta m_d \, t),
\end{align}
where $A_{\mathrm{dir}}$ and $A_{\mathrm{mix}}$ are the direct and mixing-induced CP asymmetry parameters, respectively.
To allow possible deviation from the quantum coherence, an exponential damping factor is introduced:
\begin{align}
A_\lambda(t) &= e^{-\lambda t} \left[ A_{\mathrm{dir}} \cos(\Delta m_d \, t) 
     + A_{\mathrm{mix}} \sin(\Delta m_d \, t) \right],
\end{align}
where $\lambda$, the so-called decoherence parameter, is a phenomenological parameter describing the exponential suppression of the oscillatory terms.

\subsection{Simulation Procedure}
We generate pseudo-experiments (\emph{toys}) with the following steps:
\begin{itemize}

    \item Generate the true decay times $t_{\text{true}}$ from an exponential distribution with mean $\tau_{B^0}$.
    \item Smear decay times with a Gaussian resolution of width $\sigma_t$ to obtain $t_{\text{obs}}$.
    \item Assign flavour tags $q = \pm 1$ (corresponding to $B^0$ or $\bar{B}^0$ at production).
    \item Compute the true asymmetry $A(t_{\text{obs}})$ using the chosen model and true parameter values.
    \item Accept or reject each event with probability proportional to $1 + q \, A(t_{\text{obs}})$.
    \item Bin the accepted events in decay time and compute the raw asymmetry:
    \begin{align}
        A_{\text{raw}}(t_i) = \frac{N_+(t_i) - N_-(t_i)}{N_+(t_i) + N_-(t_i)},
    \end{align}
    where $N_\pm(t_i)$ are the numbers of $q=\pm 1$ events in bin $i$.
    \item Fit $A_{\text{raw}}(t_i)$ to either the decoherence free or decoherence including model.
\end{itemize}

\subsection{Parameter Extraction}

In each toy experiment, we extracted $A_{\mathrm{dir}}$, $A_{\mathrm{mix}}$, $\Delta m_d$, and $\lambda$ when relevant. The fitting procedure is based on binned  $\chi^2$ minimization, and associated uncertainties are determined from the statistical fluctuation in each bin from each toy experiment.
\subsection{Statistical Analysis}
The distributions of the fitted parameters obtained from all pseudo-experiments are used to evaluate the statistical fluctuations, defined as 
\[
\sigma_\theta = \mathrm{std}(\hat{\theta}),
\]
where $\hat{\theta}$ denotes the fitted value of the parameter for each toy sample. To assess the performance and possible bias of the fit, pull distributions are constructed as
\[
\text{Pull} = \frac{\hat{\theta} - \theta_{\mathrm{true}}}{\sigma_{\hat{\theta}}}.
\]
For an unbiased fit with correctly estimated uncertainties, the pull distribution is expected to follow a Gaussian shape with a mean of zero and a standard deviation of one.


\subsection{Profile Likelihood for $\lambda$}
A profile likelihood scan over the decoherence parameter $\lambda$ is carried out by minimizing the negative log-likelihood with respect to the parameters $(A_{\mathrm{dir}}, A_{\mathrm{mix}}, \Delta m_d)$ for each fixed value of $\lambda$. 
The upper bound on $\lambda$ at the 95\% confidence level, denoted as $\lambda_{\mathrm{UL}}$, is obtained from the point where the likelihood difference satisfies
\[
2\,\Delta\mathrm{NLL}(\lambda_{\mathrm{UL}}) = 2.71.
\]

\section{Results and discussions}\label{Results}
The toy Monte Carlo analysis was carried out using $N_{\mathrm{toys}} = 3000$ pseudo-experiments, each consisting of $N_{\mathrm{events}} = 10^5$ simulated events. 
For each toy sample, events were generated using the true parameter values listed in Table~\ref{tab:true_params}. 
The same oscillation frequency $\Delta m_d$ and decoherence parameter $\lambda$ were assumed for all the decay modes.  

\begin{table}[h!]
\centering
\begin{tabular}{lcc}
\hline
\textbf{Decay mode} & $\boldsymbol{A_{\mathrm{dir}}^{\mathrm{true}}}$ & $\boldsymbol{\eta_f A_{\mathrm{mix}}^{\mathrm{true}}}$ \\
\hline
$B_d^0 \to J/\psi K_S$~\cite{PhysRevD.107.052008} ~~&~~ $-0.007 \pm 0.012 \pm 0.014$ ~~&~~ $0.690 \pm 0.017 \pm 0.006$ \\
$B_d^0 \to J/\psi \pi^0$~\cite{PhysRevD.107.052008} & $0.04 \pm 0.12$ & $0.86 \pm 0.14$ \\
$B_d^0 \to \pi^+ \pi^-$~\cite{LHCb:2013clb} & $-0.38 \pm 0.15$ & $-0.71 \pm 0.13$ \\
\hline
\end{tabular}

\vspace{0.2cm}
\textbf{Common parameters:} 
$\Delta m_d^{\mathrm{true}} = 0.506~\mathrm{ps}^{-1}, \quad
\lambda^{\mathrm{true}} = 0~\mathrm{ps}^{-1}$
\caption{True parameter values used for different $B_d$ decay modes. 
The oscillation frequency $\Delta m_d$ and decoherence parameter $\lambda$ are common to all modes.}
\label{tab:true_params}
\end{table}
\begin{figure}[htbp]
    \centering
     \includegraphics[width=1.0\textwidth]{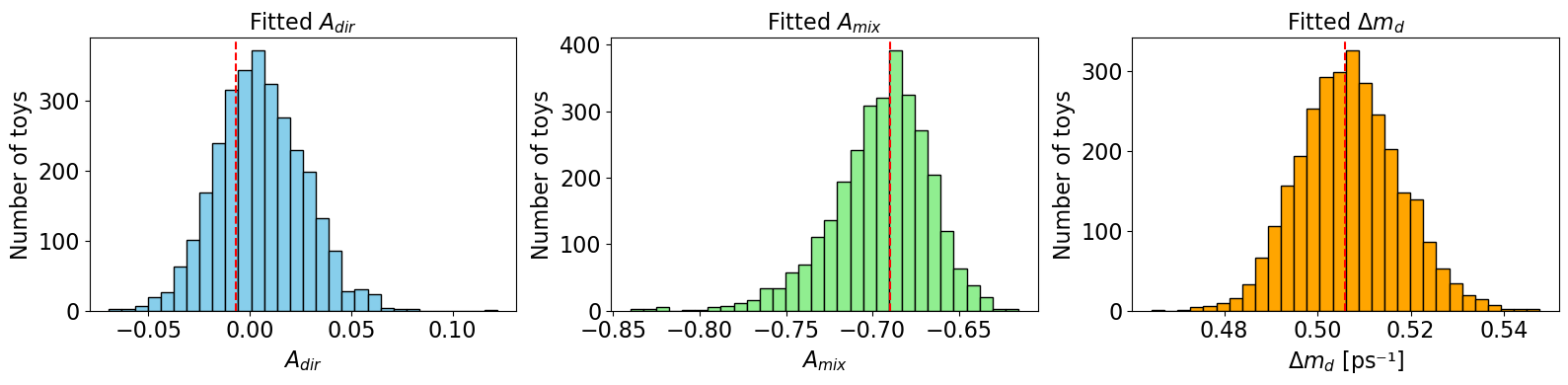}
    \includegraphics[width=1.0\textwidth]{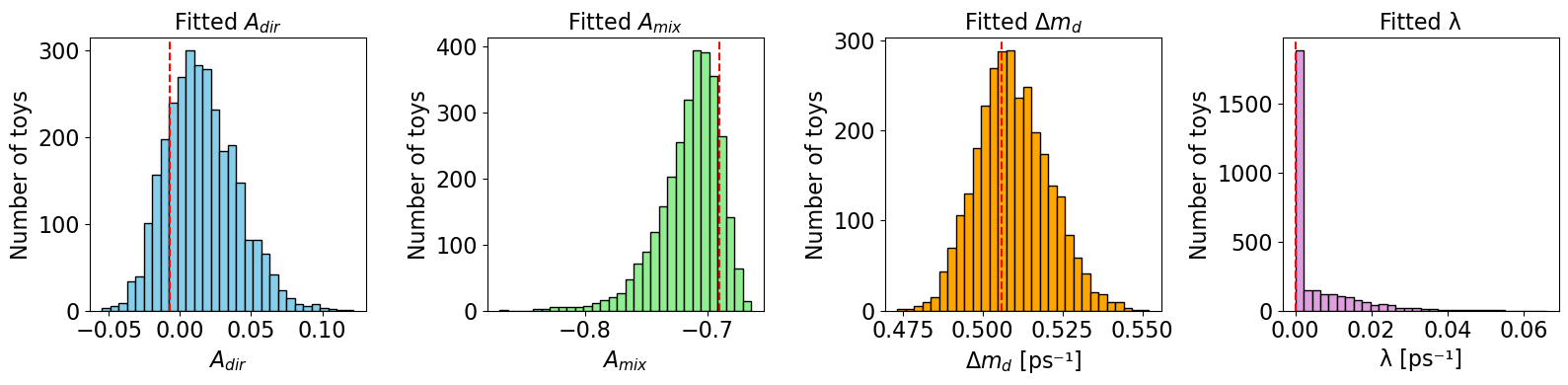}
     \caption{Distributions of fitted CP-violation parameters obtained from toy Monte Carlo samples of $B_d^0 \to J/\psi K_S$ decays (upper: without decoherence and  Lower: with decoherence.)}
    \label{fig:example1}
    \end{figure}
    \begin{figure}[htbp]
    \includegraphics[width=1.0\textwidth]{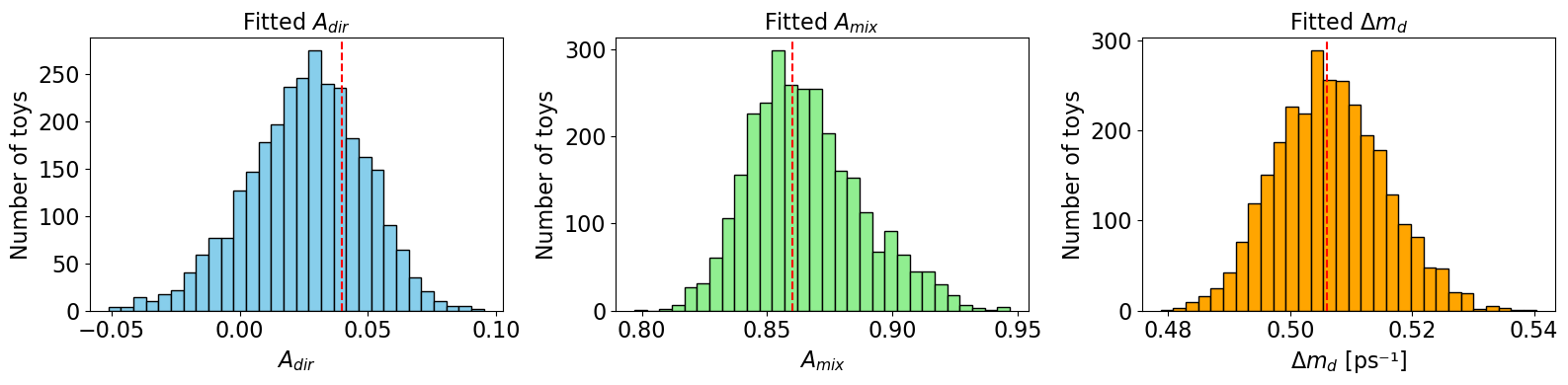}
    \includegraphics[width=1.0\textwidth]{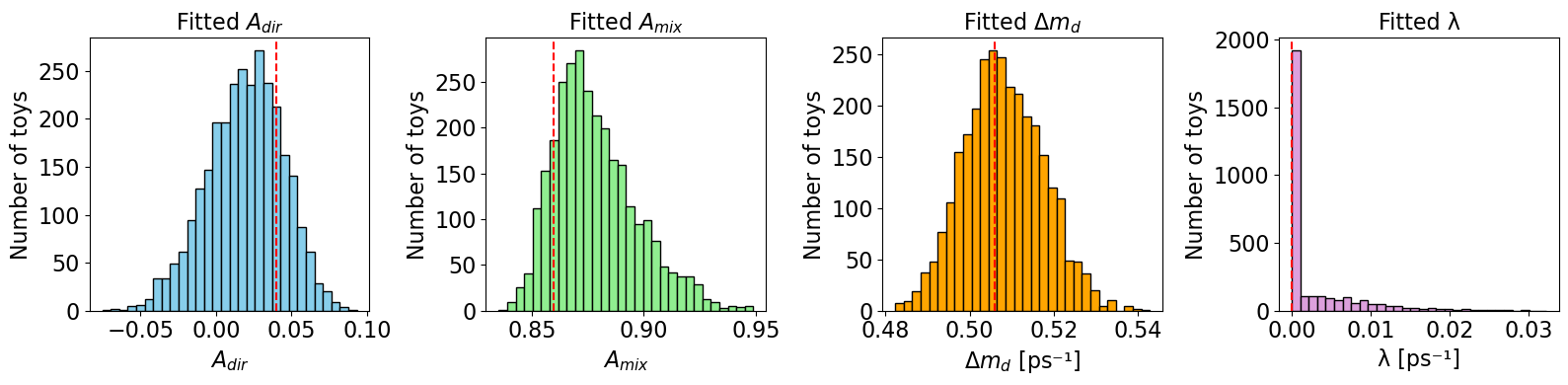}
     \caption{Distributions of fitted CP-violation parameters obtained from toy Monte Carlo samples of $B_d^0 \to J/\psi \pi^0$ decays (upper: without decoherence and  Lower: with decoherence.)}
    \label{fig:example2}
    \end{figure}
    \begin{figure}[htbp]
    \includegraphics[width=1.0\textwidth]{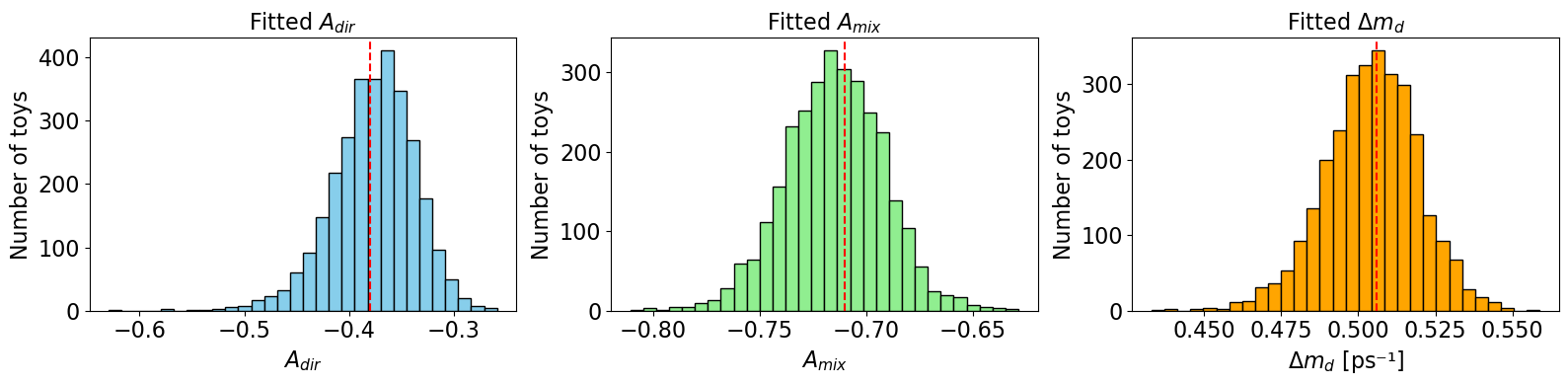} 
    \includegraphics[width=1.0\textwidth]{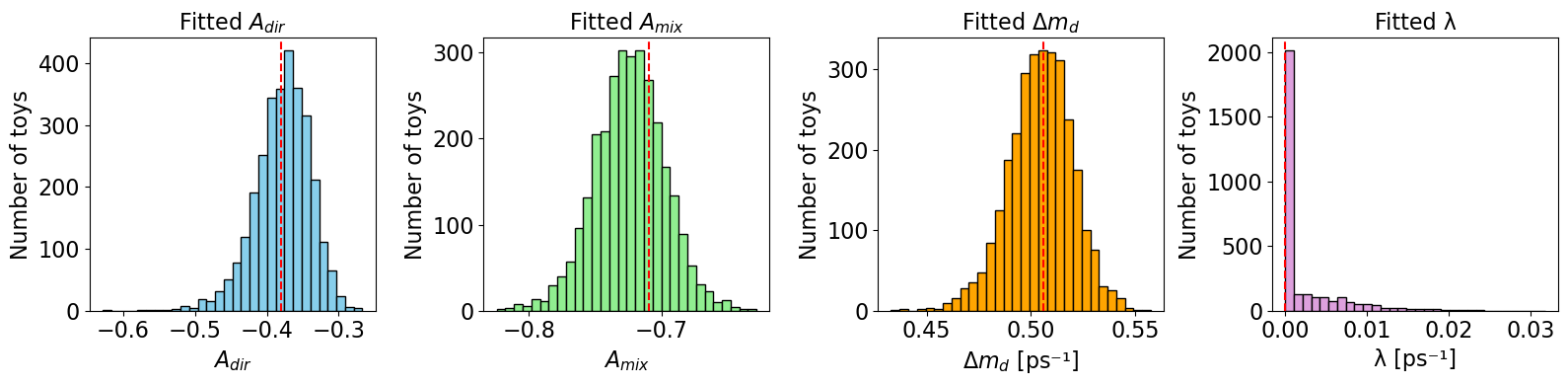}
    \caption{Distributions of fitted CP-violation parameters obtained from toy Monte Carlo samples of $B_d^0 \to \pi^+ \pi^-$ decays (upper: without decoherence and  Lower: with decoherence.)}
    \label{fig:example3}
\end{figure}
\subsection{Fitted Parameter Distributions}
The distributions of the fitted parameters $A_{\mathrm{dir}}$, $A_{\mathrm{mix}}$, and $\Delta m_d$ obtained from all toy experiments exhibiting approximately the Gaussian behavior, with their means located close to the true input values. These distributions are shown in Figs.~\ref{fig:example1},~\ref{fig:example2}, and~\ref{fig:example3} respectively for  $B_d^0 \to J/\psi K_S$,  $B_d^0 \to J/\psi \pi^0$ and  $B_d^0 \to \pi^+ \pi^-$ processes. We also summarize the fitted numerical results of three decay modes in Table~\ref{tab:results}. 
When decoherence effects are neglected, the extracted parameter values agree well with the true inputs within statistical uncertainties, demonstrating that the model provides a consistent description of the data. 
However, the presence of the decoherence parameter $\lambda$ leads to modest but noticeable shifts and broadenings in the fitted distributions for the $B_d^0 \to J/\psi K_S$ and $B_d^0 \to J/\psi \pi^0$ modes, while no such effect is observed for the $B_d^0 \to \pi^+ \pi^-$ mode, as summarized in Table~\ref{tab:results}.
 This behavior suggests that even small levels of decoherence can influence the extraction of CP-violating parameters, highlighting the importance of such effects in precision measurements.

Although the best-fit values of $\lambda$ remain compatible with zero within uncertainties, this analysis provides the first systematic exploration of decoherence effects in the non-leptonic $B_d^0$ decay modes $B_d^0 \to J/\psi K_S^0$, $B_d^0 \to J/\psi \pi^0$, and $B_d^0 \to \pi^+ \pi^-$. The resulting limits on $\lambda$ establish quantitative bounds on possible departures from the quantum coherence in the $B_d^0$ system. 

These findings reinforce the robustness of quantum mechanical coherence in $B$ meson oscillations, while also demonstrating that precision time-dependent CP analyses can serve as sensitive probes of fundamental quantum effects. With upcoming high-luminosity data from Belle~II and LHCb Upgrade, the methodology presented here can be extended to reach remarkable sensitivity to possible decoherence phenomena.

\begin{table}[h!]
\centering
\begin{tabular}{|c||c|c|c|}
\hline
\multirow{2}{*}{\textbf{Parameters}} &
\multicolumn{3}{c|}{\textbf{Decay Modes}} \\
\cline{2-4}
& $B_d^0 \to J/\psi K$ & $B_d^0 \to J/\psi \pi^0$ & $B_d^0 \to \pi^+ \pi^-$ \\
\hline\hline
\multicolumn{4}{|c|}{\textbf{Without decoherence}} \\
\hline
$A_{\text{dir}}$     & $0.0051 \pm 0.0215$  & $0.0228 \pm 0.0237$  & $-0.3755 \pm 0.0438$ \\
$\eta_f A_{\text{mix}}$     & $0.695 \pm 0.0283$  & $0.8657 \pm 0.0227$  & $-0.7163 \pm 0.0249$ \\
$\Delta m_d  (ps^{-1})$   & $0.506 \pm 0.0108$  & $0.507 \pm 0.0094$  & $0.504 \pm 0.0164$ \\
\hline
\multicolumn{4}{|c|}{\textbf{With decoherence}} \\
\hline
$A_{\text{dir}}$     & $0.0151 \pm 0.0248$  & $0.0159 \pm 0.0265$  & $-0.3771 \pm 0.0425$ \\
$\eta_fA_{\text{mix}}$     & $0.7145 \pm 0.0254$ & $0.8792 \pm 0.0190$  & $-0.7261 \pm 0.0284$ \\
$\Delta m_d (ps^{-1})$  & $0.509 \pm 0.0113$  & $0.508 \pm 0.0099$  & $0.504 \pm 0.0165$ \\
$\lambda (ps^{-1})$  & $0.0052 \pm 0.0089$  &  $0.0029 \pm 0.0052$   & $0.0024 \pm 0.0044$ \\
\hline
\end{tabular}
\caption{Comparison of theoretical predictions without and with decoherence.}
\label{tab:results}
\end{table}
\begin{figure}[htbp]
    \centering
    \includegraphics[width=1.0\textwidth]{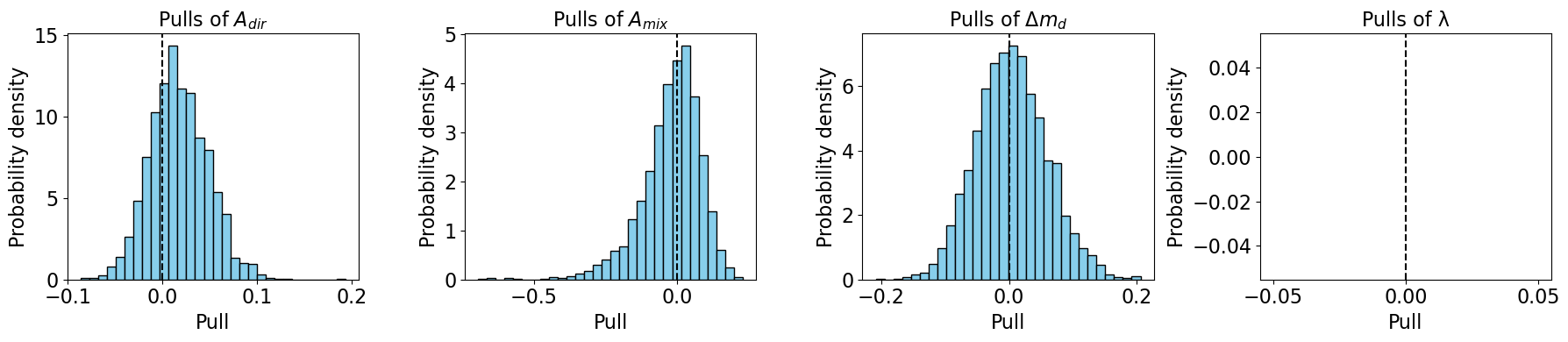}
    \includegraphics[width=1.0\textwidth]{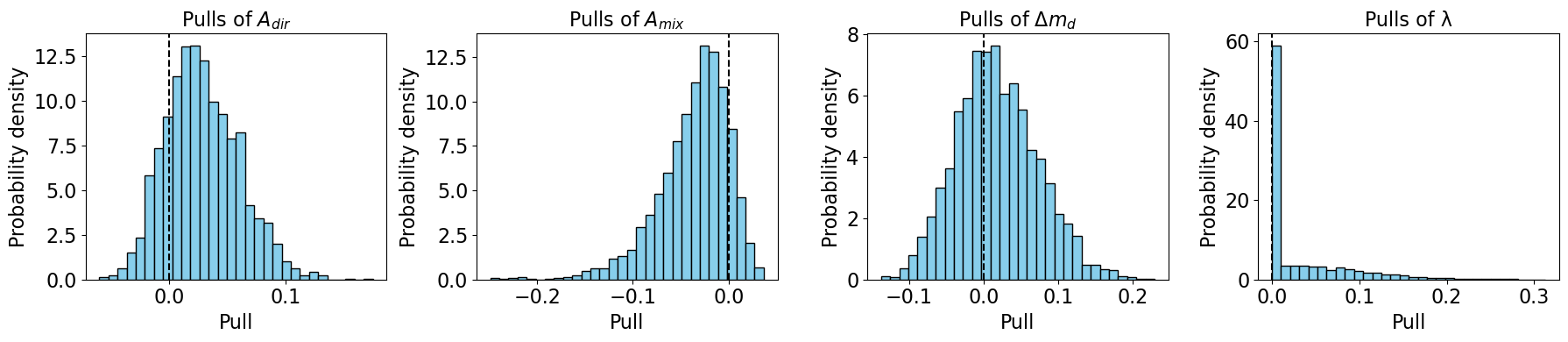} 
     \caption{Pull distributions (upper: without decoherence, lower: with decoherence) of the fitted parameters for $B_d^0 \to J/\psi K_S$ mode obtained from toy Monte Carlo studies.}
    \label{fig:example4}
\end{figure}
  \begin{figure}[htbp]
    \centering
    \includegraphics[width=1.0\textwidth]{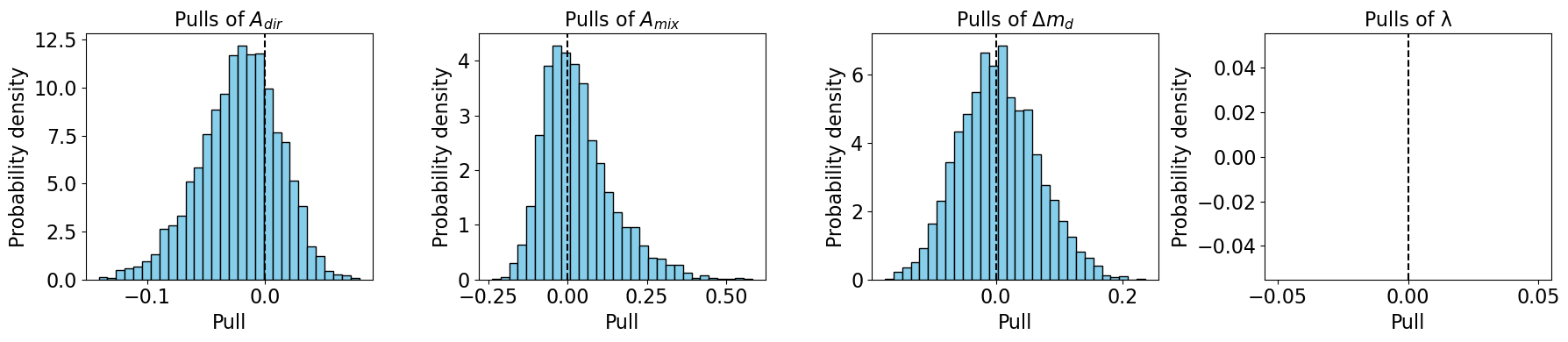} 
    \includegraphics[width=1.0\textwidth]{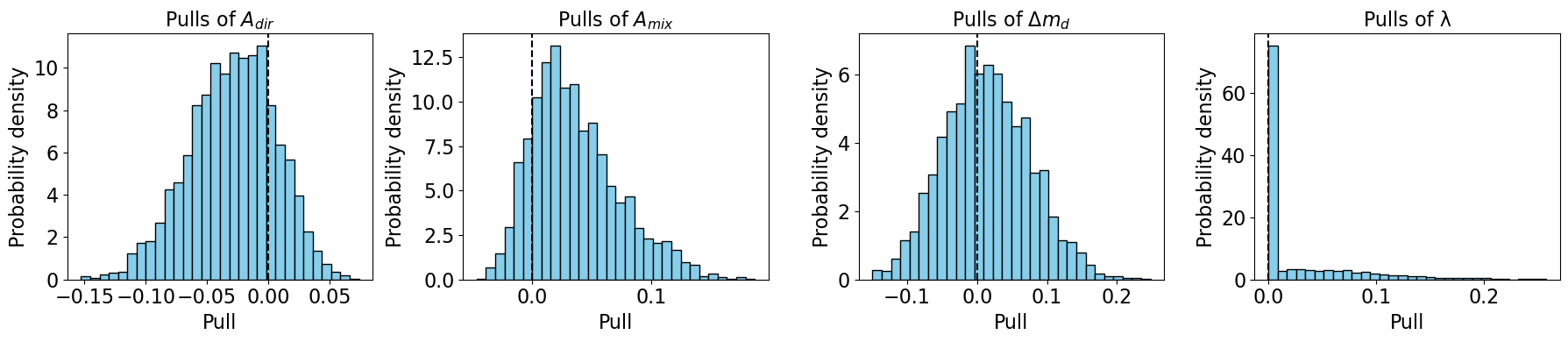} 
     \caption{Pull distributions (upper: without decoherence, lower: with decoherence) of the fitted parameters for $B_d^0 \to J/\psi \pi^0$ decays obtained from toy Monte Carlo studies.}

    \label{fig:example5}
\end{figure}
\subsection{Pull Distributions}
\begin{figure}[htbp]
    \centering
    \includegraphics[width=1.0\textwidth]{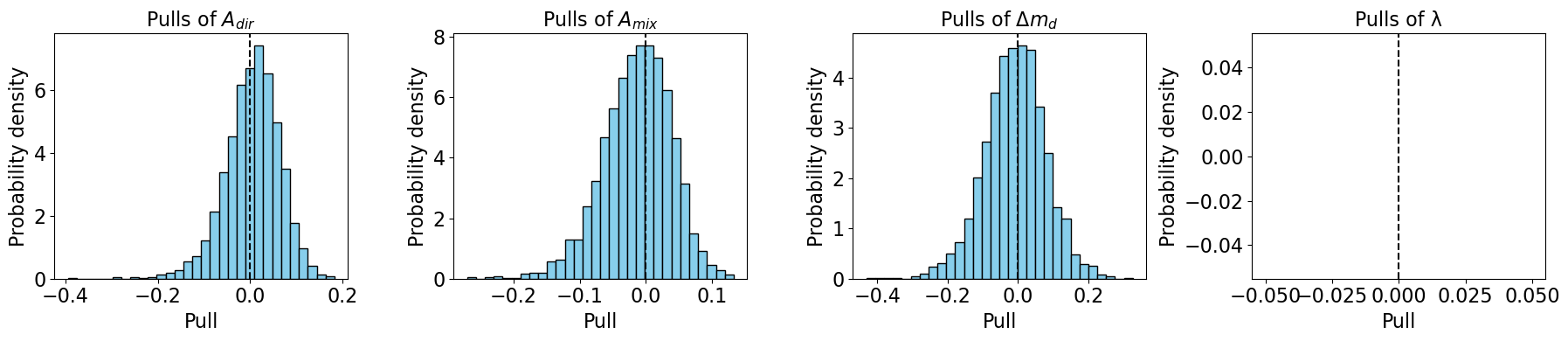} 
    \includegraphics[width=1.0\textwidth]{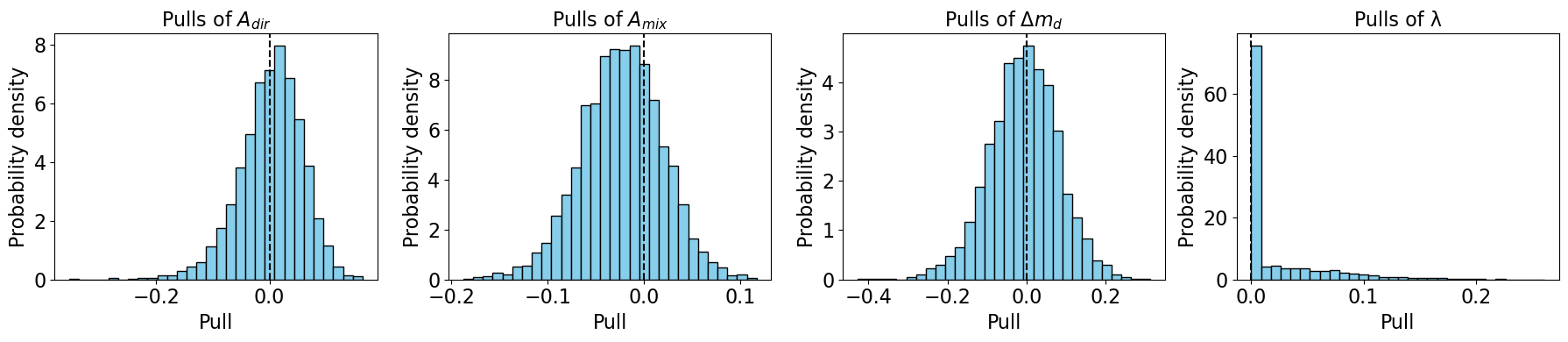} 
     \caption{Pull distributions (upper: without decoherence, lower: with decoherence) of the fitted parameters for $B_d^0 \to \pi^+ \pi^-$ decays obtained from toy Monte Carlo studies.}
    \label{fig:example6}
\end{figure}

The pull distributions corresponding to $A_{\mathrm{dir}}$, $A_{\mathrm{mix}}$, and $\Delta m_d$ exhibit Gaussian behavior, with mean values close to zero and widths compatible with unity. This agreement confirms that the uncertainties are properly estimated. The corresponding distributions for the three decay modes are illustrated in Figs.~\ref{fig:example4}–\ref{fig:example6}, and the numerical results are summarized in Table~\ref{tab:pulldecays}. 
This consistency across all three decay modes demonstrates the robustness of the fitting framework and validates the use of pseudo-experiments in modeling the underlying physics. Moreover, the fact that the pull distribution of the decoherence parameter $\lambda$ also follows the same behavior when floated in the fit further supports the stability and reliability of the fitting method. These results confirm that any deviations observed in subsequent analyses originate from the physical effects under investigation, rather than from statistical or methodological biases.

\begin{table}[h!]
\centering
\renewcommand{\arraystretch}{1.2}
\begin{tabular}{|l||c|c|c|}
\hline
\multirow{2}{*}{\textbf{Parameters}} &
\multicolumn{3}{c|}{\textbf{Decay Modes}} \\
\cline{2-4}
& $B_d^0 \!\to\! J/\psi K$ & $B_d^0 \!\to\! J/\psi \pi^0$ & $B_d^0 \!\to\! \pi^+ \pi^-$ \\
\hline\hline
\multicolumn{4}{|c|}{\textbf{Without decoherence}} \\
\hline
$A_{\text{dir}}$       & $\mu = 0.017,\; \sigma = 0.031$ & $\mu = -0.026,\; \sigma = 0.036$ & $\mu = 0.007,\; \sigma = 0.065$ \\
$A_{\text{mix}}$       & $\mu = -0.021,\; \sigma = 0.104$ & $\mu = 0.033,\; \sigma = 0.11$ & $\mu = -0.018,\; \sigma = 0.056$ \\
$\Delta m_d~$ & $\mu = 0.004,\; \sigma = 0.056$ & $\mu = 0.008,\; \sigma = 0.065$ & $\mu = -0.007,\; \sigma = 0.094$ \\
\hline
\multicolumn{4}{|c|}{\textbf{With decoherence}} \\
\hline
$A_{\text{dir}}$       & $\mu = 0.028,\; \sigma = 0.032$ & $\mu = -0.029 ,\; \sigma = 0.035$ & $\mu = 0.004,\; \sigma = 0.062$ \\
$A_{\text{mix}}$       & $\mu = -0.036,\; \sigma = 0.037$ & $\mu = 0.038,\; \sigma = 0.037$ & $\mu = -0.027,\; \sigma = 0.046$ \\
$\Delta m_d$ & $\mu = 0.018,\; \sigma = 0.065$ & $\mu = 0.018,\; \sigma = 0.65$ & $\mu = -0.006,\; \sigma = 0.095$ \\
$\lambda~$    & $\mu = 0.031,\; \sigma = 0.050$ & $\mu = 0.028,\; \sigma = 0.049$ & $\mu = 0.023,\; \sigma = 0.041$ \\
\hline
\end{tabular}
\caption{Summary of pull means and widths ($\mu$, $\sigma$) for fitted parameters in $B_d^0$ decays, comparing fits without and with decoherence.}
\label{tab:pulldecays}
\end{table}
\subsection{Profile Likelihood for $\lambda$}

\begin{figure}[htbp]
\centering
\includegraphics[width=0.45\textwidth]{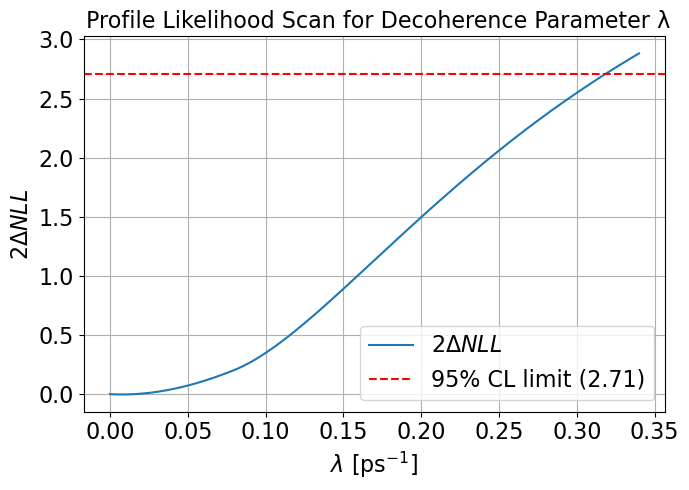} 
\includegraphics[width=0.45\textwidth]{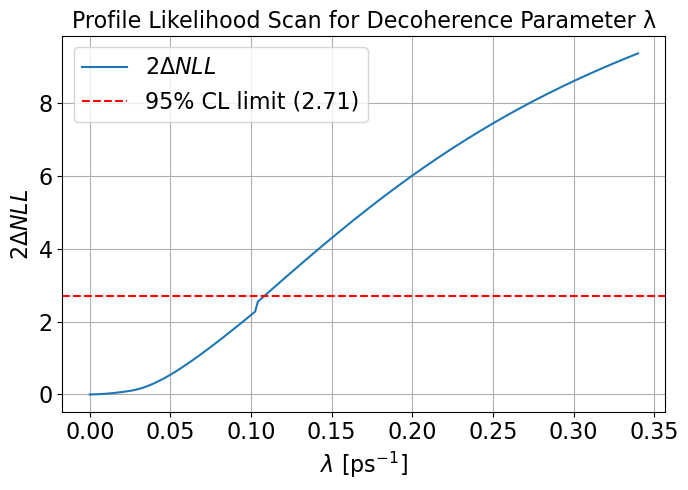} 
\includegraphics[width=0.45\textwidth]{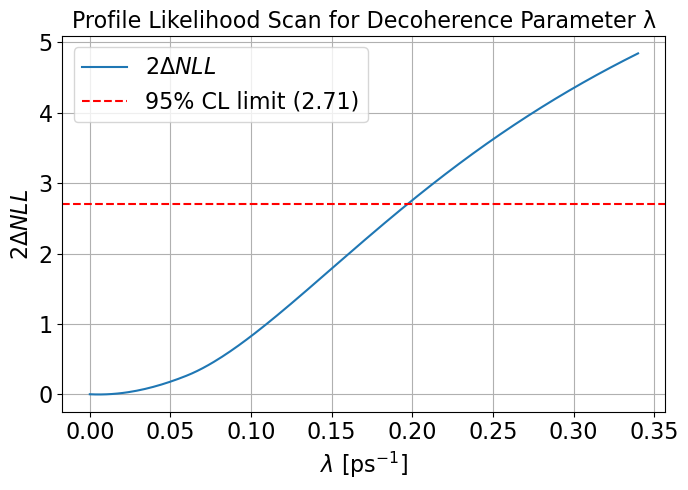} 
\caption{Profile likelihood scans for the decoherence parameter $\lambda$ in three $B_d^0$ decay modes.
Top-left: $B_d^0 \to J/\psi K_S$, top-right: $B_d^0 \to J/\psi \pi^0$, and bottom: $B_d^0 \to \pi^+ \pi^-$.
The curves illustrate the variation of $2\Delta \mathrm{NLL}$ as a function of $\lambda$, where the minima correspond to the best-fit values. The horizontal dashed line indicates the $95\%$ confidence level (2.71), used to set the upper limits on $\lambda$.}
\label{fig:example9}
\end{figure}

The profile likelihood scans, shown in Fig.~\ref{fig:example9}, exhibit minima around $\lambda = 0$ for all three decay modes, indicating mild deviation from perfect quantum coherence. The extracted $95\%$ confidence level upper limits on the decoherence parameter are $\lambda < 0.3195~\text{ps}^{-1}$ for $B_d^0 \to J/\psi K_S$, $\lambda < 0.1093~\text{ps}^{-1}$ for $B_d^0 \to J/\psi \pi^0$, and $\lambda < 0.1982~\text{ps}^{-1}$ for $B_d^0 \to \pi^+ \pi^-$, as summarized in Table~\ref{uplimits}. 

These limits demonstrate that the data are consistent with the absence of decoherence effects within the current statistical precision. The smallness of the upper limits highlights the sensitivity of our analysis to possible quantum decoherence in $B_d^0$--$\bar{B}_d^0$ mixing. With larger data samples in future experiments, these bounds can be further tightened, offering a more stringent test of the preservation of quantum coherence in neutral $B$ meson systems. 

\begin{table}[h!]
\centering
\begin{tabular}{lc}
\hline
Decay mode & $\lambda_{\mathrm{UL}}~[\mathrm{ps}^{-1}]$ \\
\hline
$B_d^0 \to J/\psi K_s$   & 0.3195 \\
$B_d^0 \to J/\psi \pi^0$ & 0.1093 \\
$B_d^0 \to \pi^+ \pi^-$  & 0.1982 \\
\hline
\end{tabular}
\caption{95\% CL upper limits on the decoherence parameter $\lambda$ obtained from profile likelihood scans for different $B_d^0$ decay modes.}
\label{uplimits}
\end{table}
\subsection{Analysis of the parameters and the observables}
\begin{figure}[htbp]
    \centering
    \begin{minipage}{0.42\textwidth}
        \centering
        \includegraphics[width=\linewidth]{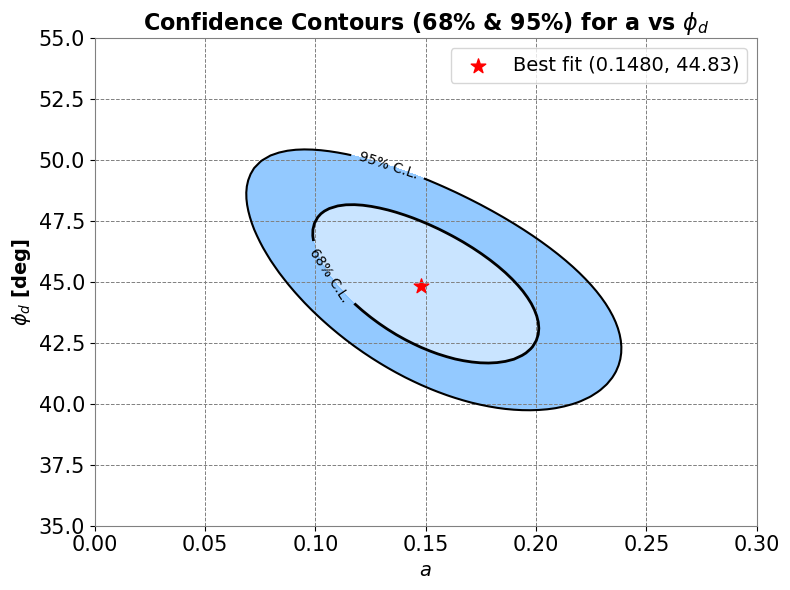}
          \includegraphics[width=\linewidth]{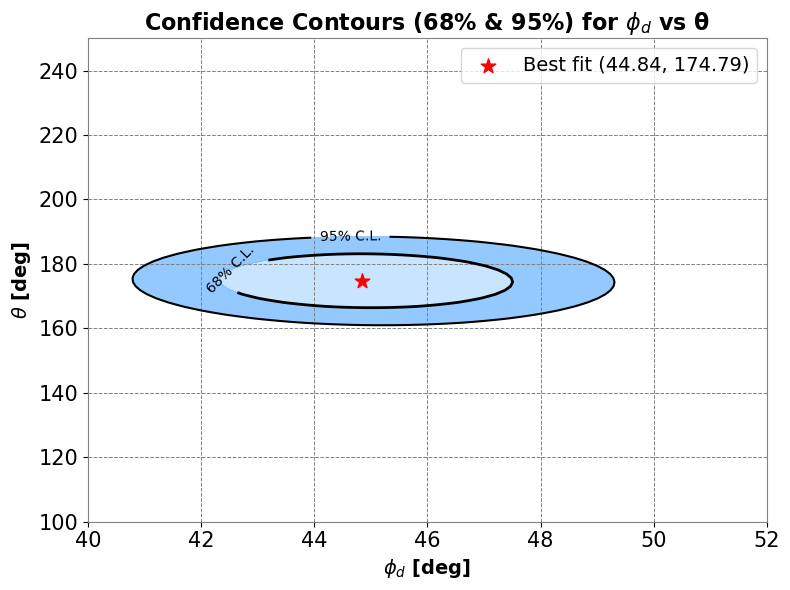}
           \includegraphics[width=\linewidth]{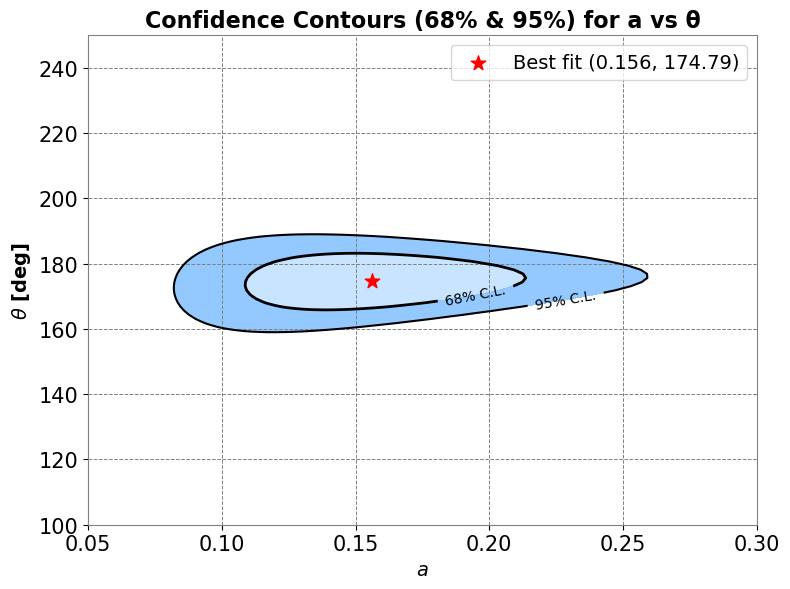}
       \caption{Correlation between the strong phase $a$, $\theta$ and the weak phase $\phi_d$ obtained from fits without decoherence for $B_d^0$ decays. The contour illustrates the allowed parameter region at  $68\%$ and $95\%$ confidence levels.}
        \label{fig:fig1}
    \end{minipage}
    \hfill
    \begin{minipage}{0.42\textwidth}
        \centering
          \includegraphics[width=\linewidth]{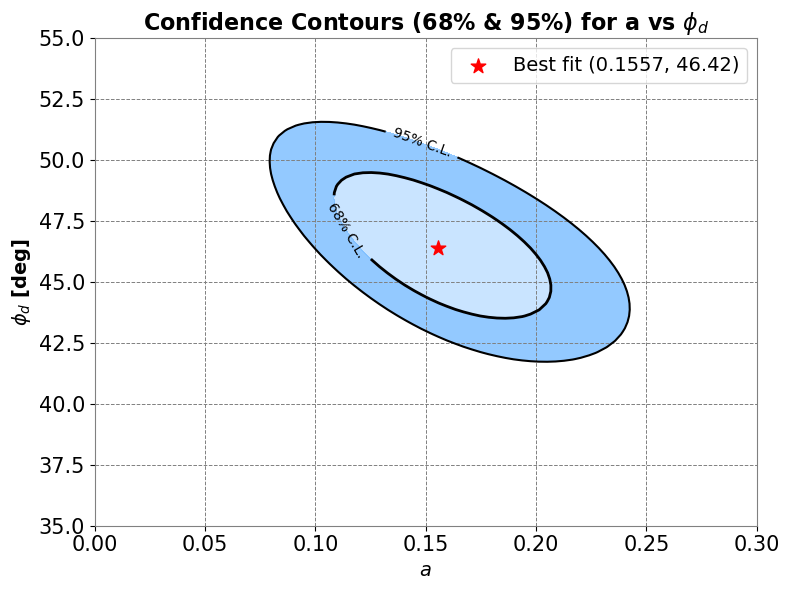}
         \includegraphics[width=\linewidth]{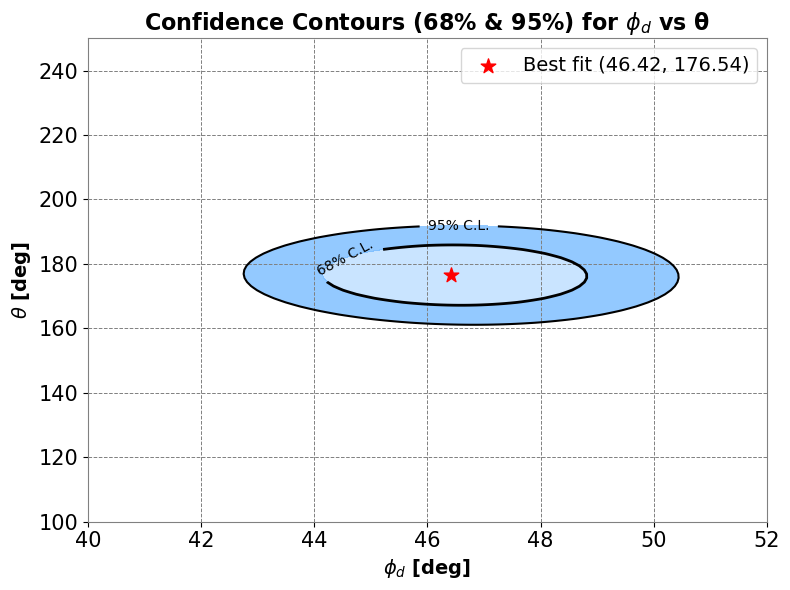}
          \includegraphics[width=\linewidth]{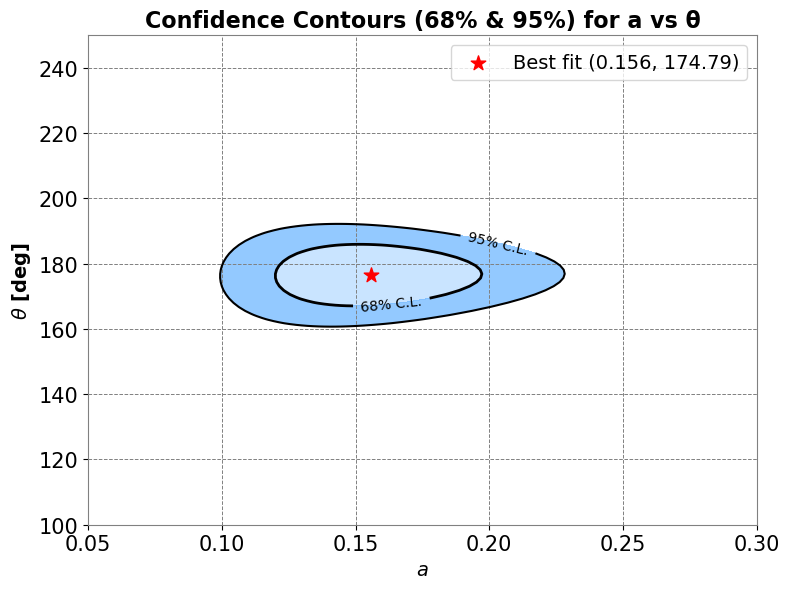}
        \caption{Correlation between the strong phase  $a$, $\theta$ and the weak phase $\phi_d$ obtained from fits with decoherence for $B_d^0$ decays. The contour illustrates the allowed parameter region at $68\%$ and $95\%$ confidence levels.}

        \label{fig:fig2}
    \end{minipage}
\end{figure}

To examine the impact of decoherence, we plotted the correlation of the penguin parameter and the CKM phase with and without decoherence in Fig.~\ref{fig:fig1}. The decoherence causes noticeable changes in the extracted penguin parameters and the CKM phase $\phi_d$ compared to the perfectly coherent case. These variations illustrate the potential of such analyses to probe small departures from quantum mechanical coherence in $B_d^0$--$\bar{B}_d^0$ oscillations. 
\begin{table}[h!]
\centering
\begin{tabular}{|c|c|c|c|}
\hline
Scenario & $a$ & $\theta$ & $\phi_d$ \\
\hline
From True value & 0.1609$\pm0.1173$ &166.22°$\pm18.08°$& 44.45°$\pm1.09°$ \\
Fit without decoherence & 0.1561$\pm0.0254$ &174.79°$\pm3.94°$ & 44.84°$\pm1.51°$ \\
Fit with decoherence & 0.1557$\pm0.0228$ & 176.54°$\pm4.37°$ & 46.42°$\pm1.39°$ \\
\hline
\end{tabular}
\caption{Values of parameters $a$, $\theta$, and $\phi_d$ for the true scenario, and fits performed with and without accounting for decoherence.}
\label{tab:scenarios}
\end{table}

The numerical results in Table~\ref{tab:scenarios} show that the fitted parameters agree well with their true input values when decoherence is neglected, confirming the reliability of the analysis. When decoherence is included, the parameter distributions become slightly broader and show small shifts, which could be tested experimentally with better statistics and improved detector sensitivity in future $B$-factory and LHCb upgrades.

As a further study, we explored the decay $B_d^0 \to \pi^- \pi^+$ and determined the value of $\alpha_{\mathrm{eff}}$ under different scenarios, which are listed in Table~\ref{tab:alpha_eff}. The estimated penguin parameters and CKM phase $\phi_d$ suggest that decoherence can distort their true values, as well as that of $\alpha_{\mathrm{eff}}$. Therefore, it is important for experimental analyses to consider the effects of decoherence when extracting precise values, even if the impact appears small.
\begin{table}[h!]
\centering
\begin{tabular}{|c|c|c|c|}
\hline
\textbf{$\alpha_{\text{eff}}$} & \textbf{From true value} & \textbf{Fit without decoherence} & \textbf{Fit with decoherence} \\
\hline
$\alpha_{\text{eff}}$ & $115.07^\circ \pm 6.68^\circ$ & $115.31^\circ \pm 1.36^\circ$ & $115.81^\circ \pm 1.57^\circ$ \\
\hline
\end{tabular}
\caption{Comparison of extracted $\alpha_{\text{eff}}$ for $B_d^0 \to \pi^+\pi^-$ from true value, without, and with decoherence.}
\label{tab:alpha_eff}
\end{table}
\begin{table}[h!]
\centering
\begin{tabular}{|c||c|c||c|c||c|c|}
\hline
\multirow{2}{*}{Decay Modes} &
\multicolumn{2}{c||}{Experimental value} &
\multicolumn{2}{c||}{Without decoherence} &
\multicolumn{2}{c|}{With decoherence} \\
\cline{2-7}
& $ A_{dir}$  & $\eta_f A_{mix}$ & $ A_{dir}$  & $\eta_f A_{mix}$ & $ A_{dir}$ & $\eta_fA_{mix}$\\
\hline
$B_d^0\to J/\psi K_s$ & -0.007$\pm$0.12 & 0.690$\pm0.17$ & 0.0051$\pm$0.0215 & 0.695$\pm $0.0283& 0.0151$\pm$0.248& 0.7145$\pm$0.254\\
\hline
$B_d^0 \to J/\psi \pi^0 $ & 0.04$\pm0.12$ &0.86$\pm$0.14& 0.0228$\pm$0.0237 & 0.8657$\pm$0.0227 & 0.0159$\pm$0.0265 & 0.8792$\pm$0.0190 \\
\hline
$B_d^0 \to \pi^+ \pi^{-}$ & -0.38$\pm0.15$ & -0.71$\pm0.13$ & -0.3755$\pm$0.438 & -0.7163$\pm$0.0249& -0.3771$\pm$0.0425 & -0.7261$\pm$ 0.0284 \\
\hline
\end{tabular}
\caption{Fitted and experimental values of $A_{\mathrm{dir}}$ and $\eta_f A_{\mathrm{mix}}$ for three $B_d^0$ decay modes, with and without decoherence.}
\label{Final}
\end{table}
To end of our analysis, we present the experimental values and the fitted spectra of the observables obtained (i) without and (ii) with decoherence for the decay modes $B_d^0 \to J/\psi K_S^0$, $B_d^0 \to J/\psi \pi^0$, and $B_d^0 \to \pi^+ \pi^-$. The comparison of the fitted parameters in both cases for the three $B_d^0$ decay modes is summarized in Table~\ref{Final}. The results show that:
\begin{itemize}
    \item  For the golden mode $B_d^0 \to J/\psi K_S$, the Standard Model already describes the data very well, leaving almost no room for decoherence effects. Introducing decoherence only causes small shifts in the central values and increases the uncertainties without improving the fit.

  \item In the flavor-partner mode $B_d^0 \to J/\psi \pi^0$, a similar behavior is observed. The fitted $\eta_f A_{\mathrm{mix}}$ value slightly increases when decoherence is included, suggesting a potential sensitivity of this mode to weak environmental interactions.
    
  \item For the hadronic decay $B_d^0 \to \pi^+ \pi^-$, both $A_{\mathrm{dir}}$ and $\eta_f A_{\mathrm{mix}}$ are in excellent agreement with the experimental averages. The decoherence effect causes only a small shift, confirming that this mode is largely unaffected at the present sensitivity level.
\end{itemize}

Overall, the fitted results reproduce the experimental observations well in the absence of decoherence. When decoherence is included, minor deviations appear, particularly in the $B_d^0 \to J/\psi \pi^0$ mode, hinting that future precision data could reveal measurable decoherence-induced effects.




\section{Conclusion}\label{conclusion}
In this work, we have investigated the possible effects of quantum decoherence in non-leptonic $B_d^0$ decays using a detailed toy Monte Carlo simulation framework. The study focused on the benchmark modes $B_d^0 \to J/\psi K_S^0$, its U-spin partner $B_d^0 \to J/\psi \pi^0$, and the complementary channel $B_d^0 \to \pi^+ \pi^-$. We generated and fitted pseudo-data to extract the time-dependent CP asymmetry parameters $A_{\mathrm{dir}}$, $A_{\mathrm{mix}}$, and $\Delta m_d$, both with and without the decoherence parameter $\lambda$.

Our results show that, in the absence of decoherence, the fitted observables reproduce the true input values within statistical uncertainties, confirming the consistency of the simulation. When decoherence effects are introduced, the fitted parameters exhibit mild shifts and broadenings, indicating that quantum decoherence could subtly affect the extraction of CP-violating observables. The pull distributions remain Gaussian, ensuring that uncertainties are properly estimated. A profile likelihood scan yields upper limits on $\lambda$ at the 95\% confidence level, showing no statistically significant deviation from perfect coherence.

Although no clear evidence for decoherence is observed within the present sensitivity, this study demonstrates a robust framework for quantifying its possible impact on flavor observables. With future high-precision data from Belle~II and the upgraded LHCb experiments, the methodology developed here can be used to place more stringent bounds on the decoherence parameter. It will also help explore potential signatures of physics beyond the Standard Model in the $B$ meson sector.

\section*{Declaration}
The analysis presented in this work is entirely based on simulated data generated under the Standard Model framework for time-dependent CP asymmetries, with an additional phenomenological parameter introduced to account for possible decoherence effects. No real experimental data were used; the results are intended purely for methodological illustration.  The introduction of the decoherence parameter $\lambda$ is motivated by the possibility that such effects could originate from physics beyond the Standard Model. A re-analysis of existing experimental data with $\lambda$ treated as a free parameter could provide a new test for deviations from standard quantum coherence in $B^0$--$\bar{B}^0$ mixing and decay. This approach would offer improved sensitivity to potential sources of new physics.

\section{Acknowledgement}
DP would like to acknowledge the support of the Prime Minister's Research Fellowship, provided by the Government of India. MKM acknowledges the financial support from IoE PDRF, University of Hyderabad.  We thank Saurabh Rai for the helpful discussion.\\

\bibliographystyle{ieeetr} 
\bibliography{Deco}   

\end{document}